\documentclass[twocolumn]{aastex63}

\newcommand\n[1][1cm]{\newline}

\shorttitle{PRASSE - The Pulsar Automated Search Script Ensemble}
\shortauthors{Gurmehar Singh, Martin Nikolov, Jake North, and Kalee Tock}

\usepackage{float} 

\begin{document}

\title{PRASSE - The Pulsar Automated Search Script Ensemble}

\correspondingauthor{Gurmehar Singh}
\email{gurmehar@gmail.com}

\author{Gurmehar Singh}
\affiliation{Stanford University Online High School \\
415 Broadway Academy Hall, Floor 2, 8853, \\
Redwood City, CA 94063, USA}
\nocollaboration{1}

\author{Martin Nikolov}
\affiliation{Stanford University Online High School \\
415 Broadway Academy Hall, Floor 2, 8853, \\
Redwood City, CA 94063, USA}
\nocollaboration{1}

\author{Jake North}
\affiliation{Stanford University Online High School \\
415 Broadway Academy Hall, Floor 2, 8853, \\
Redwood City, CA 94063, USA}
\nocollaboration{1}

\author{Kalee Tock}
\affiliation{Stanford University Online High School \\
415 Broadway Academy Hall, Floor 2, 8853, \\
Redwood City, CA 94063, USA}
\nocollaboration{1}

\begin{abstract}

The search for pulsars produces a massive amount of data which needs to be processed and analyzed. The limited speed of manual observation necessitates the involvement of large numbers of people to keep up with data collection. This paper turns to the automated alternative by examining the methodology of an algorithm built to automatically filter through processed and reduced data, which then presents the most promising data to human observers for confirmation and more complex analysis. The benefits and shortcomings of this algorithm are examined while explaining plans for future testing.

\end{abstract}

\section{Introduction} \label{sec:intro}

When the core of a relatively massive star collapses during a supernova, one of two things may occur: if the star is massive enough, the core may form a black hole, allowing nothing to escape from its gravitational pull at the event horizon. Alternatively, a slightly less massive star may have enough mass to cause the inward force of gravity to exceed that of the repulsive force created by the electron degeneracy pressure. This creates what we know as a neutron star. Some neutron stars rotate quite fast, emitting radio waves from the area around their magnetic poles. This is due to electrons being relativistically accelerated along magnetic field lines. The acceleration is greatest near the magnetic poles, which causes those areas to be the main sources of radiation. When the beam crosses into the line of sight of Earth, we see these neutron stars as pulsating sources of radiation - which we call pulsars. 
We can observe these pulsars by taking data from radio telescope and processing them in ways to amplify the relatively weak signal that reaches our telescopes. These data are often processed into what are called prepfold plots, generated by the software PRESTO. \citep{2020presto} \citep{2001ransom} \citep{2002AJ....124.1788R} Figure 1 shows an example of one of these prepfold plots, which was constructed from an observation of the pulsar J1943-1237.

\begin{figure*}[ht!]
\plotone{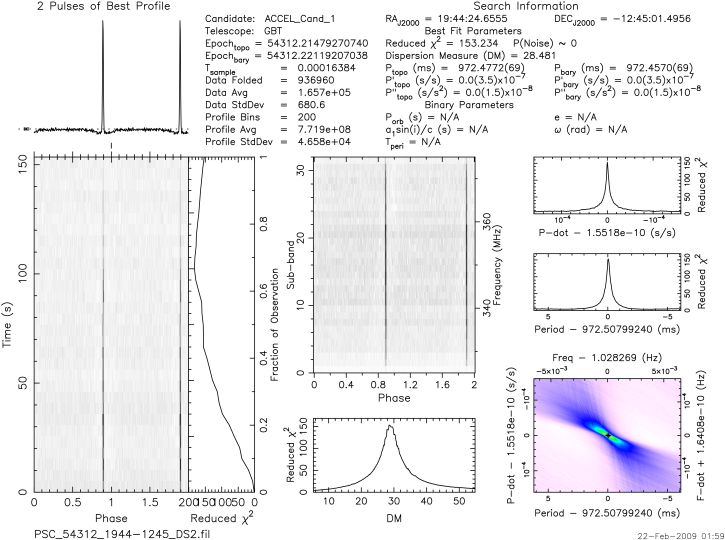}
\caption{An example prepfold plot of the pulsar J1943-1237. Note: In future plots, the information near the top and the graphs on the right side will be removed/reduced for reader convenience.}
\end{figure*}

In the center of Figure 1 is a plot that compares emission frequency to signal phase - this is aptly named the phase-frequency plot. Pulsars are highly magnetic, and electrons are accelerated near their magnetic poles, causing large amounts of synchrotron emissions to occur at their poles. This causes a radio pulse to be seen as the emission axis crosses our line of sight, and this consists of emissions at a variety of radio frequencies. For this reason, pulsar emissions show up as dark, vertical lines. The darker areas signify more intense emission, and if the lines are vertical, this shows that a broadband pulse was received. The way this clear emission is visible is only through a relatively long observation - the signal of a pulsar does not seem to stand out from background noise at first. The data are “folded” in a sense, and if done correctly, the signal of the pulsar’s pulse will become amplified. One can imagine a long sheet of paper with a graph of general pulse intensity on the vertical axis, compared to time, on the long, horizontal axis. If the correct intervals are chosen, the periodic pulse of the pulsar will become amplified and the background noise will average out. 
There are two main factors that have to be accounted for when folding these data - the first is related to the dispersion of pulses due to the ISM, called dispersion measure. Dispersion measure, or DM, is a measure of the number of free electrons present between the observer and the pulse source. Given that we have a general idea of the density of free electrons in many parts of the galaxy, given the number of free electrons present between us and the pulse source, we can estimate the distance to the source pulsar. This makes dispersion measure an effective proxy for distance. In order for the pulse to appear “lined up” as it does in Figure 1, the correct amount of “de-dispersion” must occur. A large range of dispersion measures are searched through by PRESTO, and the best candidate is chosen. When the DM is chosen correctly, the pulse appears lined up and we can see the DM value that it took to produce the graph at hand. This is represented in the plot directly beneath the phase-frequency, which compares dispersion measure to the reduced chi-squared value.
In essence, a chi-squared value is a measure of how well data fit to a pre-existing model. The closer to the model the data are, the closer to 1 the value of the chi-squared is. In this case, the pre-existing model is random noise, so random noise will have a chi-squared value of 1. As well as that, a signal will not appear to deviate much from random noise until it is de-dispersed correctly, which is evident as shown in Figure 1 - the chi-squared value peaks at a dispersion measure of around 28, which shows that the signal is strongest there. A clear peak like the one in Figure 1 is a good indication of a strong signal at a certain non-zero dispersion measure, and therefore of a pulsar. The reason for specifying non-zero dispersion measure is that if the dispersion measure is zero, we can infer that the signal did not pass through any free electrons to arrive at the receiver of the telescope, meaning it was produced inside our solar system - and on Earth - so it is certainly not a pulsar, and is very likely some form of radio frequency interference.
The second parameter for folding data correctly is rotation period - the PRESTO software searches through a large number of spin periods and folds the data at all of them, again selecting the best candidate.
In the top-left corner of the plot in Figure are the two pulses of best profile - this is a generalized representation of intensity over phase. Directly under it is the time-phase plot, which compares emission throughout phase to the observation time - this can be thought of as graphing the two pulses of best profile against time.
To clarify, in all plots regarding phase, two full phases are present.

\section{Noise and Radio Frequency Interference} \label{sec:noiseandrfi}
In the massive amount of data produced by the search for pulsars, there is a commensurately massive amount of noise and radio frequency interference - or RFI. Below are some examples of these two - the differences between pulsars and non-pulsars will become clear quite quickly. 

\begin{figure}[ht!]
\plotone{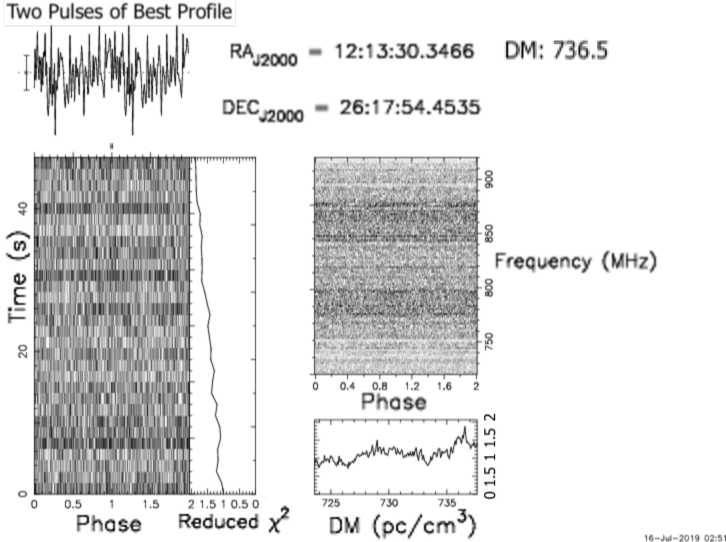}
\caption{An example plot of noise.}
\end{figure}

In Figure 2, the phase-frequency plot does not exhibit any clear vertical lines.  The appearance is analogous to that of static on a television, with no clear signal. The time-phase plot is similarly noisy, as are the two pulses of best profile. Lastly, the reduced chi-squared value has no clear peak at any dispersion measure. This is a perfect example of a plot without any clear signal - something easily marked as ”noise”.

\begin{figure}[ht!]
\plotone{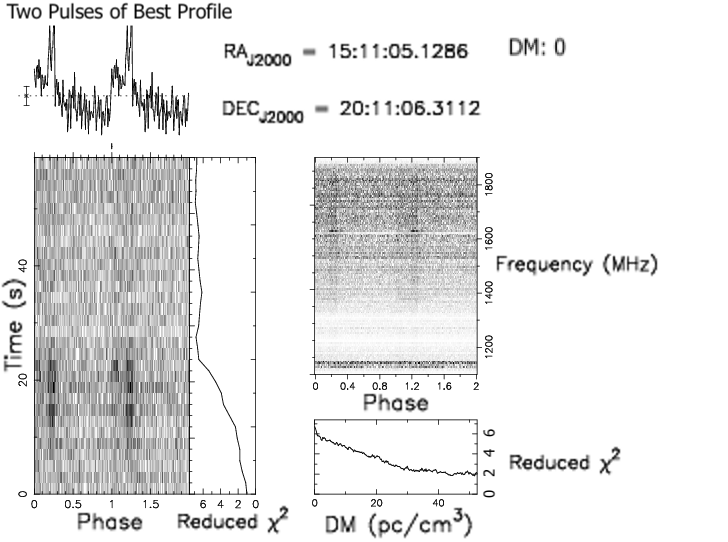}
\caption{An example plot of RFI.}
\end{figure}

RFI, on the other hand, often presents itself with certain identifiable characteristics. The first of these is a promising, pulsar-like phase-frequency plot, paired with a zero dispersion measure, which implies the signal was created on Earth. A good example of this is Figure 3. The second possibility is that the DM is calculated incorrectly - for RFI, the correct DM would be zero, but if it is calculated incorrectly, this causes the pulse to be de-dispersed when in reality, there is no de-dispersing to be done. This causes lower frequencies to be behind the higher frequencies in phase, and the signal will appear curved. A common example of this type of RFI is shown in Figure 4. The third main characteristic is narrowband emission, which is caused by a variety of man-made devices that operate only at certain frequencies. Given that pulsars are broadband sources, if narrowband emissions are seen in a plot, it can be marked off as RFI. A common example of narrowband emission is shoin Figure 5.

\begin{figure}[!ht]
\plotone{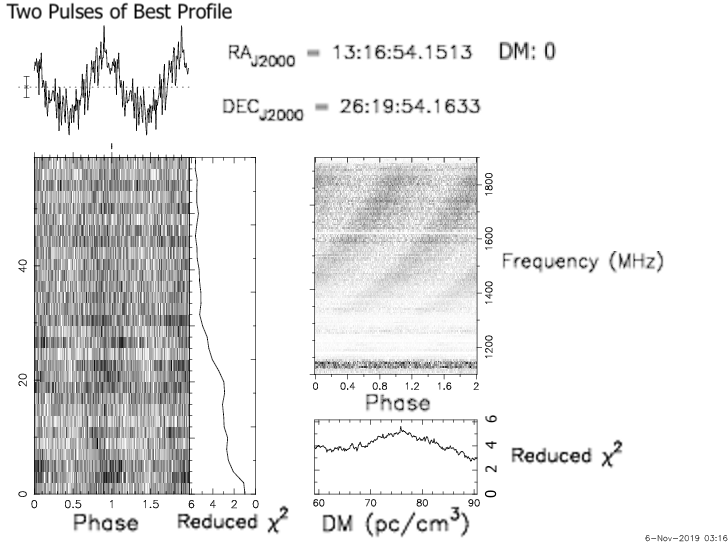}
\caption{An example plot of "curved-line RFI".}
\end{figure}

\begin{figure}[!ht]
\plotone{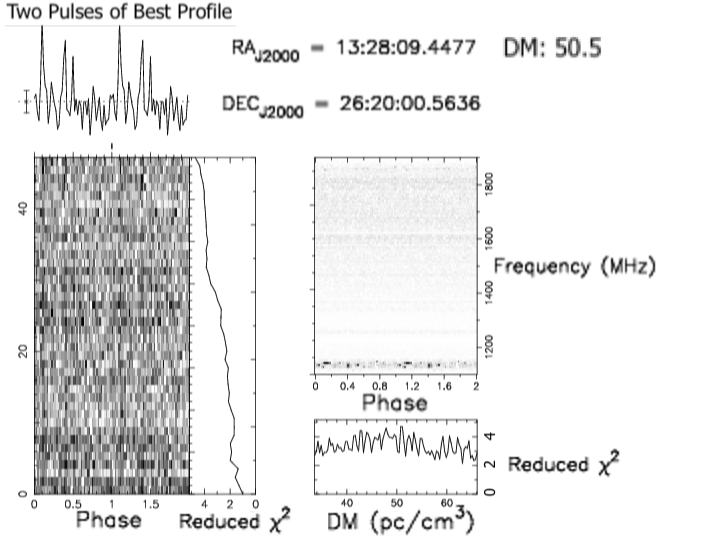}
\caption{An example plot of narrowband emission - the phase-frequency plot only indicates emission at specific frequencies.}
\end{figure}

In essence, there are three main subsets of RFI. The first is the pulsar-like, zero-DM RFI, shown in Figure 3. The second is the curved-line RFI, which is accompanied by a nonzero DM - it is crucial that the DM be nonzero in this case, since we are referring to an incorrect calculation of the DM, which we know should be zero. This is shown in Figure 4. Lastly, there is narrowband emission, which is caused by some man-made devices that operate at specific frequencies, which is shown in Figure 5.

\section{Axiswise Summation for Data Reduction} \label{sec:axisdatareduction}
The centerpiece of the prepfold plot is the phase-frequency plot. It provides almost all the information necessary to identify whether or not one is looking at a pulsar. This is because of the way it is presented - in image format. Humans are particularly good at identifying lines and patterns, and the phase-frequency plot is structured in a way that makes identifying pulsars a very straightforward process. If humans were given the raw data as a list of numbers, identifying the same patterns would prove a long and tedious task. Machines, on the other hand, specialize in being able to perform large amounts of computation, but require complex neural networks when it comes to object recognition and high-level image analysis. It follows logically from here that if images can be reduced into numerical data, it would be more convenient for machines to manipulate and analyze.

In astrophotography, it is common to stack images taken at different frequencies to amplify signal-to-noise ratio. In the phase-frequency plot, it is possible to view each frequency-band as an image taken in a separate frequency. Given that pulsars emit broadband pulses, there should be some signal in each of these "images", so if they were stacked, the result would show a significantly amplified signal-to-noise ratio. This idea became the basis of the current algorithm, and was implemented using Python and an efficient statistical analysis library, NumPy. Python was chosen for its versatility and its large range of libraries, and NumPy was chosen for its flexibility and impressive runtime.

To visualize the basic process of summation over an axis, it is helpful to think about the image as a 3-dimensional object. The first dimension would be the image height, the second its width, and the third would be for the RGB channels - as each pixel is represented by an array of three values in the format $[r, g, b]$. The first - and more trivial - summation is across this third axis. Since the phase-frequency plot is grayscale in the first place, there is no additional information provided from having three separate color channels. One can imagine laying the image down on a table and pressing it into the surface, resulting in a two-dimensional object. Then, the resulting image is compressed vertically, resulting in a list which would compare phase to the general intensity of the pulse. This result is similar to the two pulses of best profile shown in the top left corner of prepfold plots, but in this case, the result contains numerical representations about the pulses and their intensities. This summed pulse profile will be the sole piece of data analyzed by the algorithm, excepting the conditional dispersion measure reading detailed in Section 6.

\begin{figure}[!ht]
\plotone{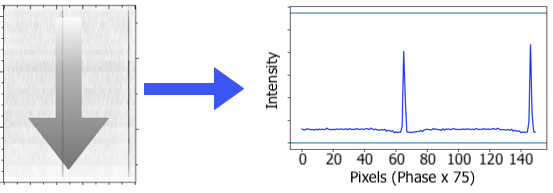}
\caption{A visualization of axiswise summation - units of intensity are relative. The image on the left is the phase-frequency plot from Figure 1, and the image on the right is the axiswise sum as described in section 3.}
\end{figure}

\section{Algorithm Specifics} \label{sec:algspec}
For those familiar with image processing and analysis, there may be some confusion as to how the summed profile was obtained in Figure 6 - since darker points in images are actually represented by lower values, it is logical to assume that a direct sum across the vertical axis would yield a profile with dips instead of peaks. This is solved by inverting the image before processing it by taking the maximum color value possible - 255 - and subtracting from it the whole image.

\centerline{$image_{new} = 255 - image_{old}$}

\subsection{Finding Significant Points}
After obtaining the summed profile in Figure 6, what remains is a list of points which represent the intensity of emission across two pulse phases - from a programming perspective, a one-dimensional array. This is quite easy to work with, and here the analysis can begin. The initial goal of this analysis is to find points in the summed profile which are significantly darker than the noise around them, which can be done by means of creating a threshold. The first step is to establish a measure of center - in this case, the median - as well as a method of detecting points that are significantly higher than the surrounding noise - in this case, the standard deviation. We can then create a threshold using the median and a multiple of the standard deviation as such:

\centerline{thresh = $med + k*SD$}

Therefore, any point in the summed profile that exceeds the value of this threshold can be considered significant.

The value for $k$ was determined from experimental statistics after tests consisting of tens of thousands of individual plots - at the time of writing, it is set at 2.7. Future testing may yield values that work more efficiently, as this algorithm is still a work in progress.

Therefore, for each point that exceeds this threshold, we can increase the "score" of the summed profile by 1. Given that the summed profile contains two phases, the score threshold would be set at 2 in the ideal scenario. However, there are certain cases, due to images sometimes being off by a few pixels at one time or another that both pulses may not be contained in the cropped-out section of the phase-frequency plot. For this reason, the score threshold is set at 1. Creating a dynamic system to accurately crop out the phase-frequency plot based on the image itself and not on predetermined coordinates is a plan for future work.

\subsection{Dynamic Thresholding}
One problem with a static threshold for the entire summed profile is that it fails to account for a systematic increase in baseline intensity in the region of a peak. In other words, a peak in emission can present itself alongside an increase in the noise it is surrounded by. This makes a static threshold more vulnerable to marking random noise and fluctuations as significant. One way to counteract this problem is to create a "dynamic" threshold, which would change for every data point. In essence, rather than taking the median and standard deviation of the entire profile, they are taken for $n$ number of points to the left and right of the data point in question. This better ensures that the peak is actually darker than the surrounding "area" and not a random fluctuation in noise. Further testing of many thousands of plots showed that the most effective value was $n=30$. For reference, the summed profile typically contains 150 individual data points, meaning one phase consists of 75 points. This means at $n=30$, the threshold is using 80\% of the phase to determine the threshold for significance.

\begin{figure}[!ht]
\plotone{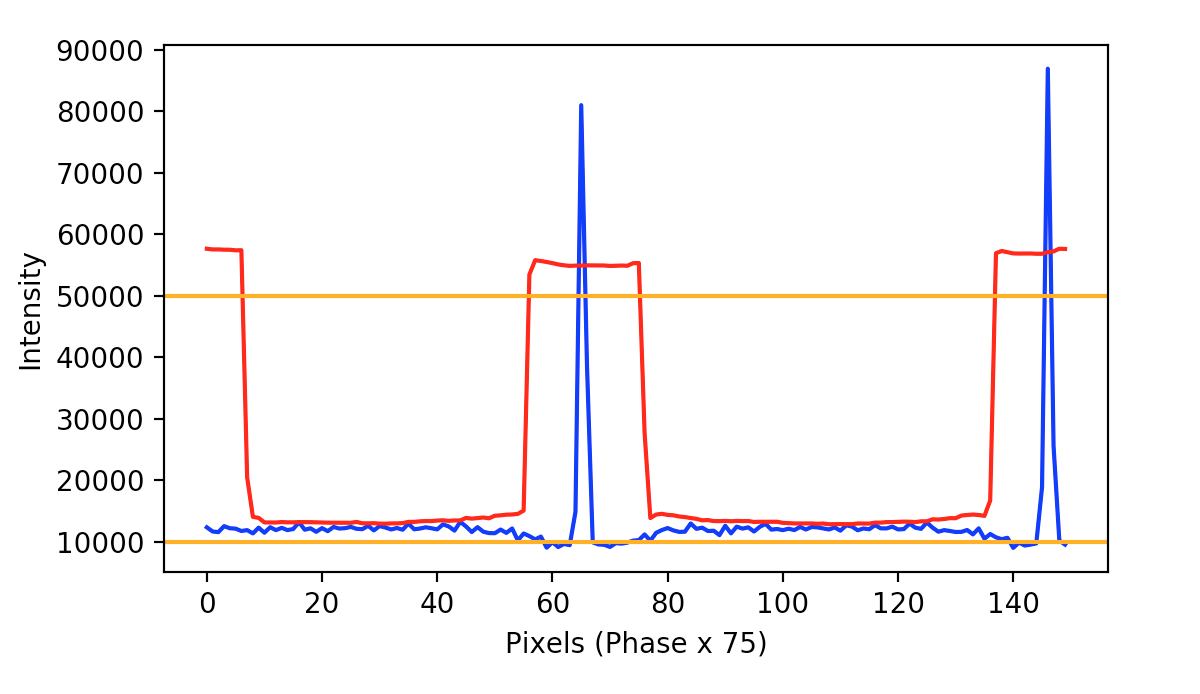}
\caption{A visualization of the dynamic threshold - the red line - and the intensity - the blue. The two orange lines act as additional thresholds, and will be explained in further detail in Section 4.3.}
\end{figure}

\subsection{Maximum and Minimum Override}
From Figure 7, the dynamic threshold seems oddly strict or "high" for a pulse. The reason for the increase in the threshold is actually due to the pulse itself. Since the dynamic threshold is calculated at each point using the surrounding \textit{n} points, that includes the pulse, which is much higher than the rest of the baseline noise. However, pulses that tend to be very pronounced and clear like the one seen in Figures 1, 5, and 6, also tend to be surrounded by very little noise, so the pulses can stand out. This, of course, is a generalization, and therefore this dynamic threshold is not perfect. One way to compensate for this is to implement so-called "maximum override" and "minimum override" parameters.

To understand the "maximum override" and "minimum override", we must take a deeper look at axiswise summation. When summing the image both across the RGB axis and the vertical axis, the traditional 0 to 255 range of RGB values changes. The first summation combines these three color channels, so the range of values becomes in effect three times larger. Given that the initial maximum value is 255, after summing three color channels together, all of the values in the resulting array are contained within 0 and $255*3$, which comes out to between 0 and 765. Then, after the summation across the vertical axis - resulting in the summed profile - the range of values increases again. This time, the upper bound will increase by a factor of the height of the image - numerically, 0 to $765 * height_{img}$. Therefore, if pulsars can be said to have some "minimum pulse intensity", most pulsars should have a point in their summed profiles which fall above this certain value in the range $[0, 765*height_{img}]$. Therefore, if the summed profile exceeds this value at any point, the plot will automatically be marked as significant or of interest. The optimal value for this parameter was determined to be 50,000 after testing on over 350 individual sample prepfold plots of pulsars. For perspective, the current number of known pulsars is approximately 3,000. This optimization test also consisted of minimizing the false-positive rate while keeping the true-positive at 100\% within the test set.

The "minimum override" ends up doing the exact opposite - it is a lower limit intended to easily filter out plots in which a large majority of the frequency channels received little to no signal. If the entirety of the summed profile lies below a certain value in the range $[0, 765*height_{img}]$, then we can be confident that the signal is either extremely weak and of no interest, or that it is narrowband and limited to very few frequency channels - explaining the low value - in which case it is RFI and also of no interest. Therefore, if the entirety of the summed profile lies below this minimum, then it can be marked as insignificant and filtered out. Again, through extensive testing, this value was determined to be 10,000.\n

The maximum and minimum overrides can be seen in orange in Figure 7.

\subsection{Notes on the Dynamic Threshold}
The dynamic threshold is a double-edged sword - with a very low value for n, the false positive rate drops to its lowest, ensuring that very little noise gets through the filter - but this may miss some weaker pulsars. Conversely, a higher value of n tends to correspond to a higher false positive rate, but also ensures that more pulsars are caught. 
The problem with the dynamic threshold is that it can cause pulses to not be detected due to an unnecessarily increased mean around the pulse itself. However, with a large enough value of n, it is possible to strike a balance between the benefits of the static and dynamic thresholds. Initial testing shows that with a value of \textit{n = 30}, the dynamic threshold performs marginally better than a static threshold at sorting out noise and maintaining the detection rate listed in Section 5. The reason for this discrepancy is currently unknown, and will be the subject of a more detailed study in the future.

\section{Experimental Statistics} \label{sec:expstat}
These statistics are based off of repeated testing with tens of thousands of pieces of data. The true positive rate comes from a test set of 124 individual pulsar plots, separate from the roughly 350 used to fine-tune the algorithm.

True Positive Rate: 100\%

False Positive Rate: ~5\% $\pm$ 1.5\%

Plots Tested and Sorted: ~100,000

The classifications of the testing plots were all known.
The false positive rate is also not a major concern for the algorithm and its development. Reducing the amount of data by roughly 95\% is already an immense difference from the thousands of plots of noise that need to be looked at manually. The overarching goal is simply to not miss real pulsars, so this algorithm is designed to be more "lenient" when evaluating candidates. The false positive rate is also highly dependent on the value of \textit{n}, the parameter regarding the dynamic threshold.

\section{Dispersion Measure Reading} \label{sec:dmread}
If a pulsar candidate is detected through the phase-frequency plot, the next step would be to check the peak dispersion measure. The current method relies on the image format of these prepfold plots being standard, and the dispersion measure being listed at a certain set of coordinates in the image file - see the written information at the top of Figure 1. The main algorithm then runs a second script to check the dispersion measure of the plot using Tesseract, an image-to-text library for Python. The dispersion measure reading is cropped out of the image, and the size of the image is increased to make conversion more accurate. Once the image is converted to text, the numerical value itself is extracted through string parsing. If the resulting dispersion measure is less than 2, the plot is sent to the RFI folder. In the case of a reading error, the image to text conversion is stopped and the plot will be marked as significant as a safeguard. Future development includes a more reliable system to extract DM from the written information on the plot without having to rely on standard image formatting.

\section{Shortcomings and Flaws} \label{sec:flaws}
The algorithm is obviously not perfect - rather than being an arbiter of "pulsar or not pulsar", it works more effectively as a filter so that only the more significant and interesting pieces of data are shown to human viewers. It is also possible that the chi-squared value for the prepfold plot - as seen in Figure 1, directly above the listed dispersion measure in the information near the top - could also provide a measure of how "significant" a plot is and whether or not it might be a pulsar or just noise.
After testing on over 100,000 individual plots, roughly 97\% of all non-pulsar plots tend to be filtered out. There are certain types of plots that make it through the filter, and improvements are being made and tested to continually increase the filter accuracy. The majority of non-pulsar plots that can make it through is "curved RFI", an example of which is seen in Figure 4. If the lines are dark enough and the curve is steep enough, they may appear as pulsar-like peaks in the summed profile. Given that this is a result of incorrect dispersion measure calculation, the nonzero dispersion measure combined with the apparent pulsar-like profile make a promising candidate, so the plot makes its way through the filter. The implementation of a convolutional neural network in the future will likely help in solving this problem.

\section{Pulsar Detections} \label{sec:pulsardetections}
Throughout the course of the testing and use of this algorithm, various pulsars have been detected. One of the first detected pulsars was J1946+2611 - a previously-known pulsar. One of the flagged plots is shown in Figure 8.

\begin{figure}[!htbp]
\plotone{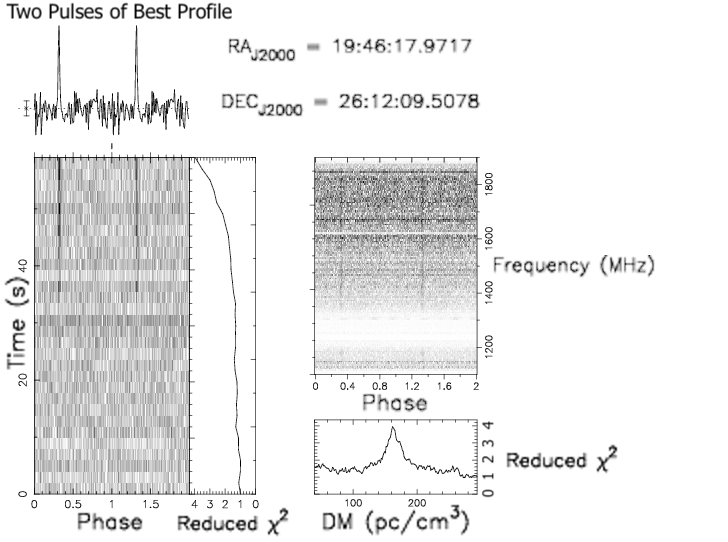}
\caption{A flagged plot of J1946+2611.}
\end{figure}

Recently, while filtering through over 30,000 individual plots, the program helped co-discover a new pulsar candidate by flagging two significant plots, which were later identified by several human observers. This candidate is quite promising and is in the process of being confirmed at Green Bank Observatory - see Figure 9 and Figure 10.

\begin{figure}[!htbp]
\plotone{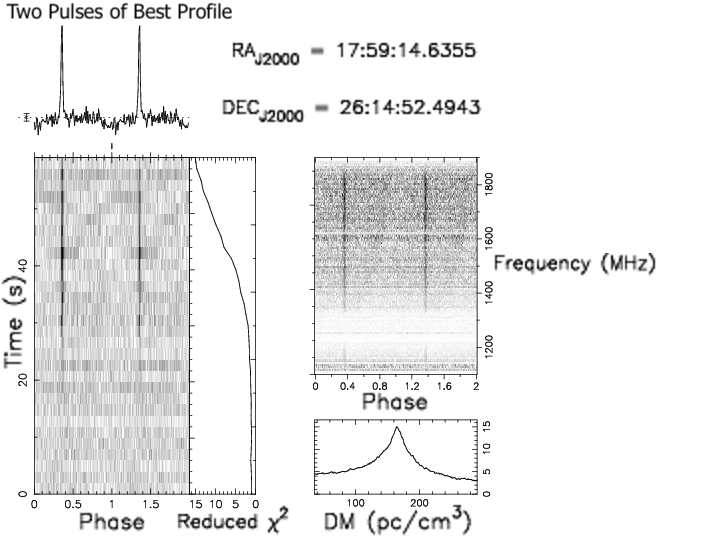}
\caption{A flagged plot of the new pulsar candidate.}
\end{figure}

\begin{figure}[!htbp]
\plotone{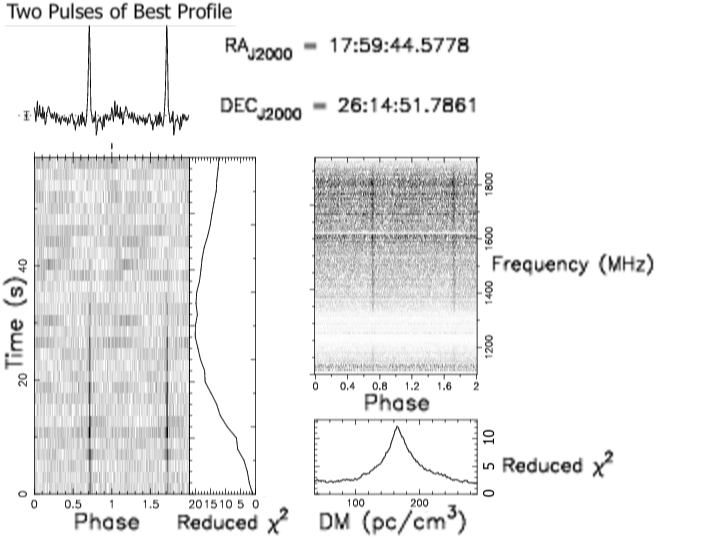}
\caption{A flagged plot of the new pulsar candidate.}
\end{figure}

\section{Final Information}\label{sec:ending}
The source code for this algorithm can be found at https://github.com/gsingh-0-0-1/prasse, along with test data, version history, and consistent updates.

\acknowledgments
We thank the Pulsar Search Collaboratory (PSC) \citep{2010AEdRv...9a0106R} for access to data which was used to test this algorithm. The discovery-centered approach of the PSC \citep{2013ApJ...768...85R} allowed us to write and structure this algorithm in the first place. The new pulsar candidate was also co-discovered from data obtained through the Pulsar Search Collaboratory.

We would also like to thank Dr. Maura McLaughlin of West Virginia University for advice and feedback on how to improve the algorithm, as well as providing a test set of over 150 pulsars for an earlier version of the program. 

\newpage
  
\bibliography{FinalDraft}{}
\bibliographystyle{aasjournal}

\end{document}